\documentclass{aa} 

\usepackage{graphicx}
\usepackage{txfonts}
\usepackage{hyperref}
\hypersetup{
    colorlinks=true,
    linkcolor=blue,
    urlcolor=magenta,
}
\usepackage{listings}

\newcommand{\Gaia}{{\it Gaia }}
\newcommand{\gspspec}{GSP-Spec}

\newcommand{\T}{$T_{\rm eff}$}
\newcommand{\g}{log($g$)}

\newcommand{\meta}{[M/H]}
\newcommand{\alphaFe}{[$\alpha$/Fe]}

\newcommand{\FeH}{[Fe/H]}
\newcommand{\CaFe}{[Ca/Fe]}

\newcommand{\MgFe}{[Mg/Fe]}

\titlerunning{Double RGB and RC sequences of disc stars with \textit{Gaia} \gspspec}
\begin{document} 
\definecolor{dkgreen}{rgb}{0,0.6,0}
\definecolor{gray}{rgb}{0.5,0.5,0.5}
\definecolor{mauve}{rgb}{0.58,0,0.82}

  \title{Double red giant branch and red clump features of Galactic disc stellar populations with \textit{Gaia} \gspspec}


    \author{Alejandra Recio-Blanco
          \inst{1}
           \and
           P. de Laverny\inst{1}
           \and
           P. A. Palicio\inst{1}
           \and
           S. Cassisi  \inst{2,3}
           \and
           A. Pietrinferni  \inst{2}
           \and
           N. Lagarde \inst{4}
           \and
           C. Navarrete\inst{1} 
           }
   \authorrunning{Recio-Blanco et al.}  \institute{\inst{1}\textit{Universit\'e C\^ote d'Azur, Observatoire de la C\^ote d'Azur, CNRS, Laboratoire Lagrange, Bd de l'Observatoire, CS 34229, 06304 Nice cedex 4, France\\
   \email{alejandra.recio-blanco@oca.eu}\\
               }
    \inst{2}\textit{INAF – Osservatorio Astronomico di Abruzzo, Via M. Maggini s/n, I-64100 Teramo, Italy}\\
    \inst{3}\textit{INFN, Sezione di Pisa, Largo Pontecorvo 3, 56127 Pisa, Italy}\\
    \inst{4}\textit{Laboratoire d'Astrophysique de Bordeaux, Universit\'e Bordeaux, CNRS, B18N, All\'ee Geoffroy Saint-Hilaire, 33615 Pessac, France
    }
}
   \date{Received ; accepted }

 
  \abstract
   {The bimodality of the Milky Way disc, in the form of a thick short disc and a thinner more radially extended one, encrypts the complex internal evolution of our Galaxy and its interaction with the
environment.}
   {To disentangle the different competing physical processes at play in Galactic evolution, a detailed chrono-chemical-kinematical and dynamical characterisation of the disc bimodality is necessary, including high number statistics.}
   {Here, we make use of an extremely precise sub-sample of the \textit{Gaia} DR3 \gspspec\ catalogue of stellar chemo-physical parameters. The selected database is composed of 408~800 stars with a median uncertainty of 10~K, 0.03, and 0.01~dex in \T, \g\  and \meta, respectively. }
   {The stellar parameter precision allows us to break the age-metallicity degeneracy of disc stars. For the first time, the disc bimodality in the Kiel diagram of giant stars is observed, getting rid of interstellar absortion issues. This bimodality produces double red giant branch sequences and red clump features for mono-metallicity populations. A comparison with BaSTI isochrones allows us to demonstrate that an age gap is needed to explain the evolutionary sequence separation, in agreement with previous age-metallicity relations obtained using sub-giant stars. A bimodal distribution in the stellar mass-\alphaFe\ plane is observed at constant metallicity. Finally, a selection of stars with \meta=0.45$\pm$0.03~dex shows that the most metal-rich population in the Milky Way disc presents an important proportion of stars with ages in the range of 5-13~Gyr, in agreement with previous literature findings. This old, extremely metal-rich population is possibly a mix of migrated stars from the internal Galactic regions, and old disc stars formed before the last major merger of the Milky Way.}
   {The \textit{Gaia} \gspspec\ Kiel diagrams of disc mono-abundance stellar populations reveal a complex, non-linear age-metallicity relation crafted by internal and external processes of Galactic evolution. Their detailed analysis opens new opportunities to reconstruct the puzzle of the Milky Way disc bimodality.}


   \maketitle
%

\section{Introduction}
The structure of the Milky Way disc has been known to be bimodal since the seminal works of \cite{Yosii82} and \cite{GilmoreReid}. It is composed of a thick, more primitive stellar population and a thin, more radially extended component containing stars, gas, and dust. The disc bimodality encrypts crucial information about the physical processes of our Galaxy formation, as it is the result of more than 10~Gyr of evolution. It emerges from the complex interplay of internal evolutionary processes and the interaction of our Milky Way with its ecosystem. As a consequence, a variety of physical mechanisms are invoked to explain the spatial, kinematical, dynamical, and chemical properties of the thick and the thin discs, and their evolution with time. Early turbulence \citep{Bournaud09}, radial migration \citep[e.g][]{SchonrichBinney,Minchev2010,Trick2019, Nikos2023}, gas infall \citep[eg][]{EmanueleGaia}, galaxy mergers \citep[e.g.][]{Abadi2003, Renaud21}, and disc marginal stabilty \citep{Park21} can be cited among the numerous studied processes and their related literature.

From an observational point of view, characterising the disc bimodality implies precise data and high number statistics. For this reason, the thick-thin disc transition has been at the core of the observational strategy of all recent spectroscopic surveys of the Milky Way. In particular, the existent correlations between chemical abundances and kinematical and/or dynamical properties have been explored inside the solar neighbourhood, using HARPS and GALAH data,  among others \citep[e.g.][]{Vardan2013, PabloDisc, Michael2020} and outside the solar neighbourhood, using for instance \textit{Gaia}-ESO Survey, APOGEE, GALAH, or \textit{Gaia} \gspspec\ data \citep[e.g.][]{GESDisc, Michael2015, Buder2021, PVP2023}.  Thick disc stellar populations are kinematically hotter than thin disc ones. 
Additional evidence of disc bimodality comes from chemical abundance ratios. In particular, the thick and thin disc populations define two parallel sequences in the \alphaFe\ abundance versus metallicity plane, with thick disc stars presenting higher \alphaFe\ ratios than thin disc stars of similar \meta, with a gap in \alphaFe\ , whose detailed properties indeed depend on the specific $\alpha$-element taken into consideration\citep{Nikos2023}. The two sequences seem to merge in the super-solar metallicity regime, although this fact is observationally challenging to confirm due to the complex analysis of metal-rich stellar spectra \citep{PabloNormalization}.
Moreover, the temporal evolution of the disc chemical properties suggests the existence of two distinct phases in the age-metallicity relation \citep[e.g.][]{MishaHarps, MichaelGES, PabloDisc, Nadege21, Miglio2021, XiangRix2022}.  In the metal-intermediate regime, a gap of about 2-5~Gyr between the mono-metallicity populations of the thick and the thin discs seems to be present. Conversely, a gap in metallicity of about 0.5~dex at the start of the thin disc formation 7-9~Gyr ago is detected. This metallicity gap is progressively reduced, and seems to disappear 4-5~Gyr ago. The relative proportion of thin to thick disc stars is linked to the different structural properties of the two discs, with the thick disc having a shorter scale-length and a larger scale-height than the thin one.

To disentangle the Galactic disc dichotomy, and in particular its temporal evolution, the study of large samples of stars in a large spatial volume is mandatory. To this purpose, giant stars are ideal tracers of disc populations up to very large distances. However, their use is hampered by a severe age-metallicity degeneracy of their physical properties (in particular, colour, and therefore effective temperature), and non-negligible uncertainties on interstellar absorption in the Galactic plane. In this letter, we use an extremely precise sample of stellar chemo-physical parameters (cf. Sect.~\ref{sect:data}) 
from the \textit{Gaia} DR3 \citep{DR3Vallenari}  \gspspec\ catalogue \citep{GSPspec2023} to break the age-metallicity degeneracy of disc stellar populations. New constraints on disc bimodality using the \T\ versus \g\ diagram are presented in  Sect.~\ref{sect:DoubleRGBs}. Finally, the age distribution of extremely metal-rich stars in the disc is explored in Sect.~\ref{sect:VeryMetalRich}. The conclusions of this letter are summarized in Sect.~\ref{sect:Conclu}.
\begin{figure}[t]
\centering
\includegraphics[width=0.5\textwidth]{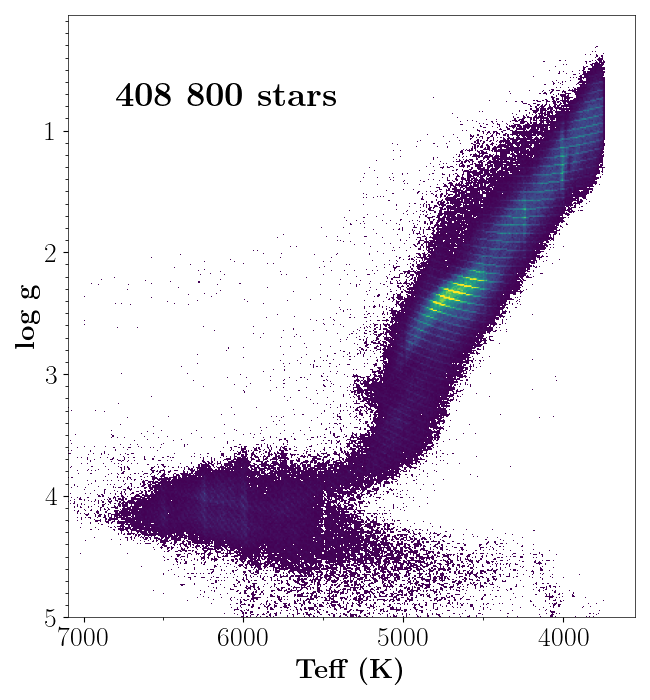}
\caption{Distribution of the selected high-precision DR3 \gspspec\ dataset in the logarithm of surface gravity vs effective temperature plane.}
\label{fig:Kiel}
\end{figure}

\begin{figure}[t]
\centering
\includegraphics[width=0.45\textwidth]{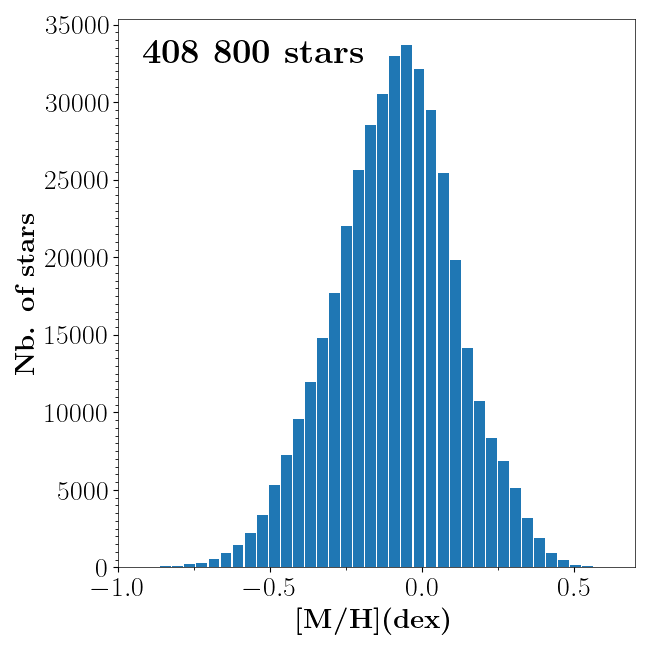}
\caption{Metallicity distribution sample of the selected high-precision DR3 \gspspec\ dataset.}
\label{fig:Mdf}
\end{figure}

\begin{figure}[t]
\centering
\includegraphics[width=0.5\textwidth]{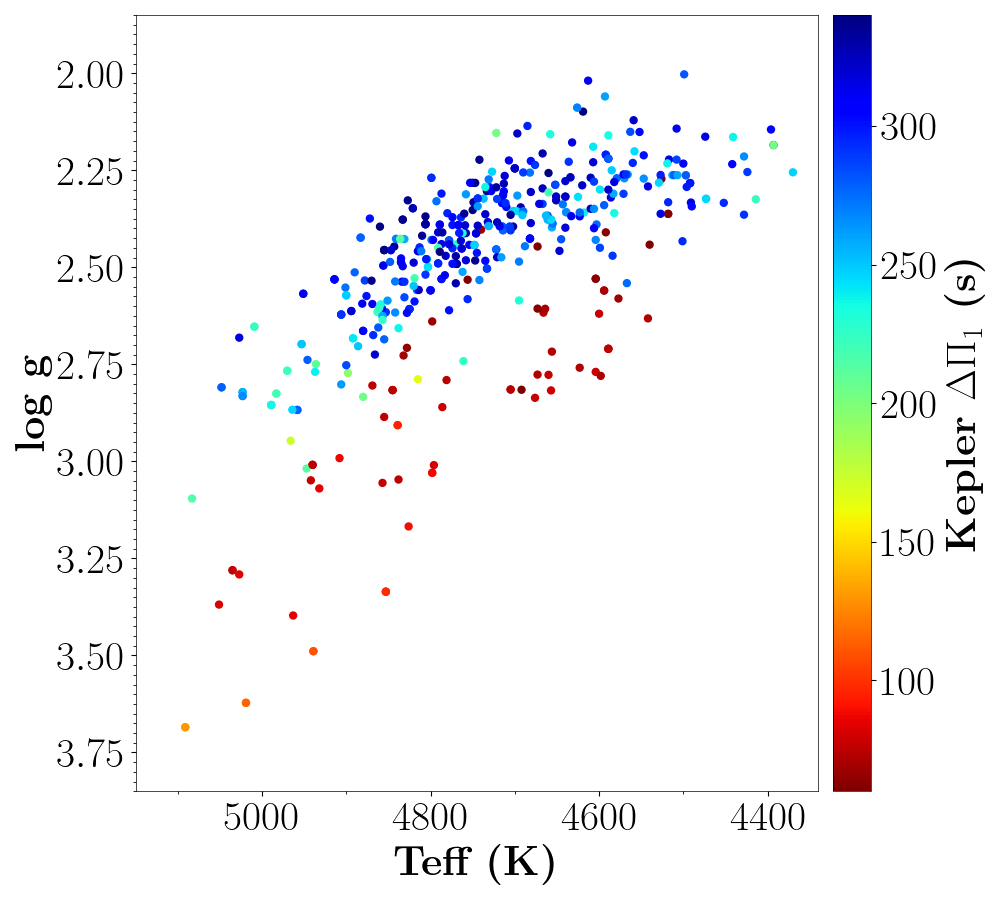}
\caption{DR3 \gspspec\ logarithm of surface gravity vs effective temperature for sources observed by the Kepler mission and included in our high-precision sample. The period spacing, $\Delta\Pi_{1}$, derived from Kepler data (not used by the \textit{Gaia} \gspspec\ module) is illustrated as a colour code, confirming the \gspspec\ parametrisation of RGB and RC giants.}
\label{fig:KielKepler}
\end{figure}

\begin{figure}[t]
\centering
\includegraphics[width=0.5\textwidth]{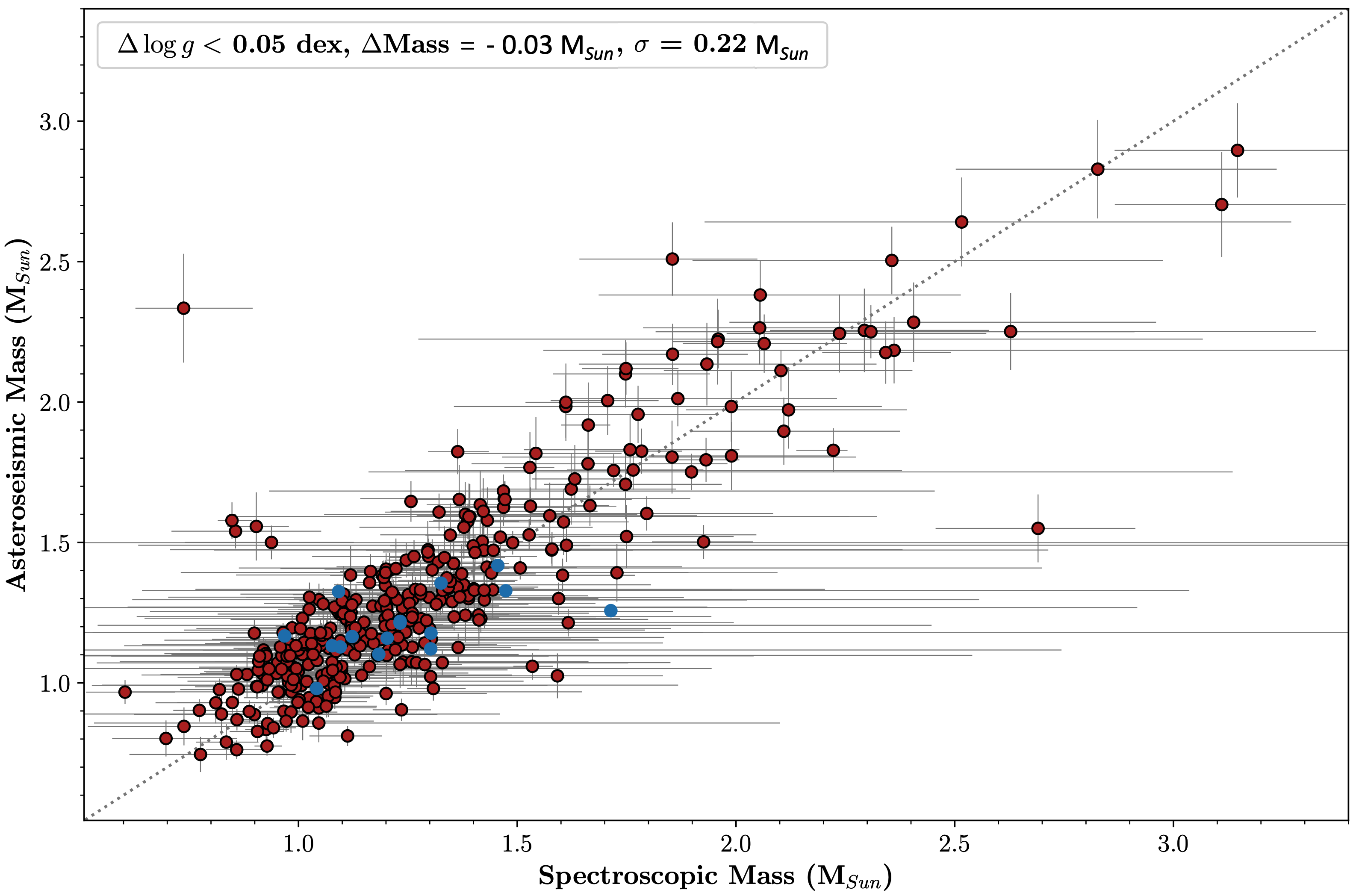}
\caption{ Comparison of our stellar spectroscopic masses derived from \gspspec\ parameters and \textit{Gaia} data with APOKASC masses of giants (red points) and dwarfs (blue points) from \cite{ApokascGiants} and \cite{ApokascDwarfs}, respectively. The selection is restricted to sources for which the agreement in \g\ is better than 0.05~dex between \gspspec\ and APOKASC, as the spectroscopic masses are sensitive to the stellar surface gravity.}
\label{fig:APOCASKMasses}
\end{figure}

\section{A high-precision DR3 \gspspec\ dataset}
\label{sect:data}
This work makes use of the stellar physical parameters and chemical abundances derived from the \Gaia RVS spectra by the MatisseGauguin analysis procedure \citep{Matisse, GSPspec2016, GauguinAlbert} of the \gspspec\ module \citep{GSPspec2023}. The data are
available through the {\it astrophysical\_parameters} table of the \Gaia DR3 archive. In particular, we employed the stellar effective temperature \T\ (\verb|teff_gspspec|), the surface gravity \g\ (\verb|logg_gspspec|), the metallicity \meta\ (\verb|mh_gspspec|), and the abundance of $\alpha$-elements \alphaFe\ and magnesium \MgFe\ with respect to iron  (\verb|alphafe_gspspec| and  \verb|mgfe_gspspec|, respectively). The prescriptions recommended by \cite{GSPspec2023} in Tables~3 and 4 have been used to calibrate \g, \meta\, \alphaFe\ , and \MgFe\ as a function of \T\ (see Appendix~\ref{App:Calibs}).
It is important to note that the \gspspec\ DR3 \meta\ abundances are indeed indicators of the \FeH\ abundance, as iron lines dominate the non-$\alpha$ element features used by \gspspec\ for the \verb|mh_gspspec| estimations. As a consequence,  \meta\ abundances can be directly compared to \FeH\ values of chemical evolution models or stellar isochrones.

To select high-precision parameters  (see Appendix~\ref{App:Query}), our working sample contains only stars with a metallicity uncertainty defined as (\verb|mh_gspspec_upper|-\verb|mh_gspspec_lower|)/2, lower than 0.025~dex, and a signal-to-noise (\verb|rv_expected_sig_to_noise|) higher than 100. Moreover, to ensure the quality of the parametrisation, we imposed the following filters in the \verb|flags_gspspec| chain \citep[cf.][Table 2]{GSPspec2023}: the \verb|vbroadT,G,M|, \verb|vradT,G,M| and the \verb|KMgiantPar| flags have been set to 0,  and the \verb|extrapol| flag to 0 or 1. Finally, to clean for the remaining less reliable solutions, we selected stars not cooler than 3750~K and with a goodness-of-fit parameter, \verb|logchisq_gspspec|<-3.7. The final sample contains 408,800 stars with a median uncertainty of 10~K, 0.03 
and 0.01~dex in \T, \g\ , and \meta\ , respectively. The signal-to-noise of the corresponding RVS spectra is distributed between 100 and 2823 with a median at 144. The $G$ magnitude is distributed between 3.47 and 12.44~mag with a median at 8.8~mag. The stars are located between 6.7~kpc to 10~kpc from the Galactic centre and within 1~kpc from the Galactic plane. 
Figures~\ref{fig:Kiel} and \ref{fig:Mdf} show the Kiel diagram and the metallicity distribution function of the final selection, respectively. It mainly contains F-type, G-type, and K-type stars with disc-like metallicities. The lack of metal-poor halo stars is essentially due to the imposed filters in metallicity uncertainty, but also to the RVS selection function for DR3 high signal-to-noise data.

 In addition to the already published DR3 \gspspec\ data, stellar luminosities and masses were derived 
using the corresponding \gspspec\ spectroscopic
atmospheric parameters, \textit{Gaia} photometry, and astrometry  (see de Laverny, in preparation, for a detailed description). 
First of all, the E(Bp-Rp) extinction was estimated comparing the observed
\textit{Gaia} (Bp-Rp) colour to the theoretical one, derived from \cite{Casagrande21} and using the \gspspec\ parameters. We then computed the bolometric correction and stellar luminosity,
adopting the geometric distances from \cite{Coryn21}. The stellar mass was finally derived from this luminosity, using the corresponding
effective temperature and surface gravity. The associated errors were estimated by
propagating
the uncertainties on the parameters, photometry, and distances, thanks to
Monte-Carlo realisations.
Finally, we have used the kinematical and orbital parameters of the selected stars computed by \cite{Pedro2023a} to complement the spectroscopic \gspspec\ chemo-physical ones.

\subsection{Cross-match with asteroseismic data}
To test the precision of the selected \textit{Gaia} \gspspec\ data and their derived parameters, we performed a cross-match with sources observed by the NASA Kepler mission and the APOCASK project.
Firstly, Fig.~\ref{fig:KielKepler} shows the \gspspec\ Kiel diagram of 468 stars with Kepler observations and included in our high-precision \gspspec\ sample. The Kepler period spacing parameter, $\Delta\Pi_{1}$, derived by \cite{Mosser12}, \cite{Mosser14}, and \cite{Vrard16} is used as a colour code in the figure. The $\Delta\Pi_{1}$ parameter is a reliable indicator of the stellar evolutionary stage. As is shown in Fig.~\ref{fig:KielKepler}, although the \gspspec\ module does not use any asteroseismology constraints, the Kepler $\Delta\Pi_{1}$ is perfectly coherent with the \gspspec parametrisation that is capable of separating the stars  in the red giant branch (RGB; $\Delta\Pi_{1}\lesssim$ 150 s) from those in the red clump (RC; $\Delta\Pi_{1}\gtrsim$ 150 s).

Secondly, Fig.\ref{fig:APOCASKMasses} shows a comparison of the spectroscopic masses derived in this work and those provided by \cite{ApokascGiants} and \cite{ApokascDwarfs} combining Kepler asteroseismology and APOGEE high-resolution spectroscopy \citep[see][for the most recent release]{APOGEE_DR16}, as part of the APOKASC project. For coherence, we have restricted our cross-match to the sources presenting an agreement in the derived surface gravity better than 0.05~dex, as the spectroscopic masses are very sensitive to the assumed \g. Our derived masses show a very good agreement with respect to APOKASC, with a bias of only -0.03~M$_{\sun}$ and a dispersion of 0.22~M$_{\sun}$.

\begin{figure*}[h]
\centering
\includegraphics[width=0.9\textwidth]{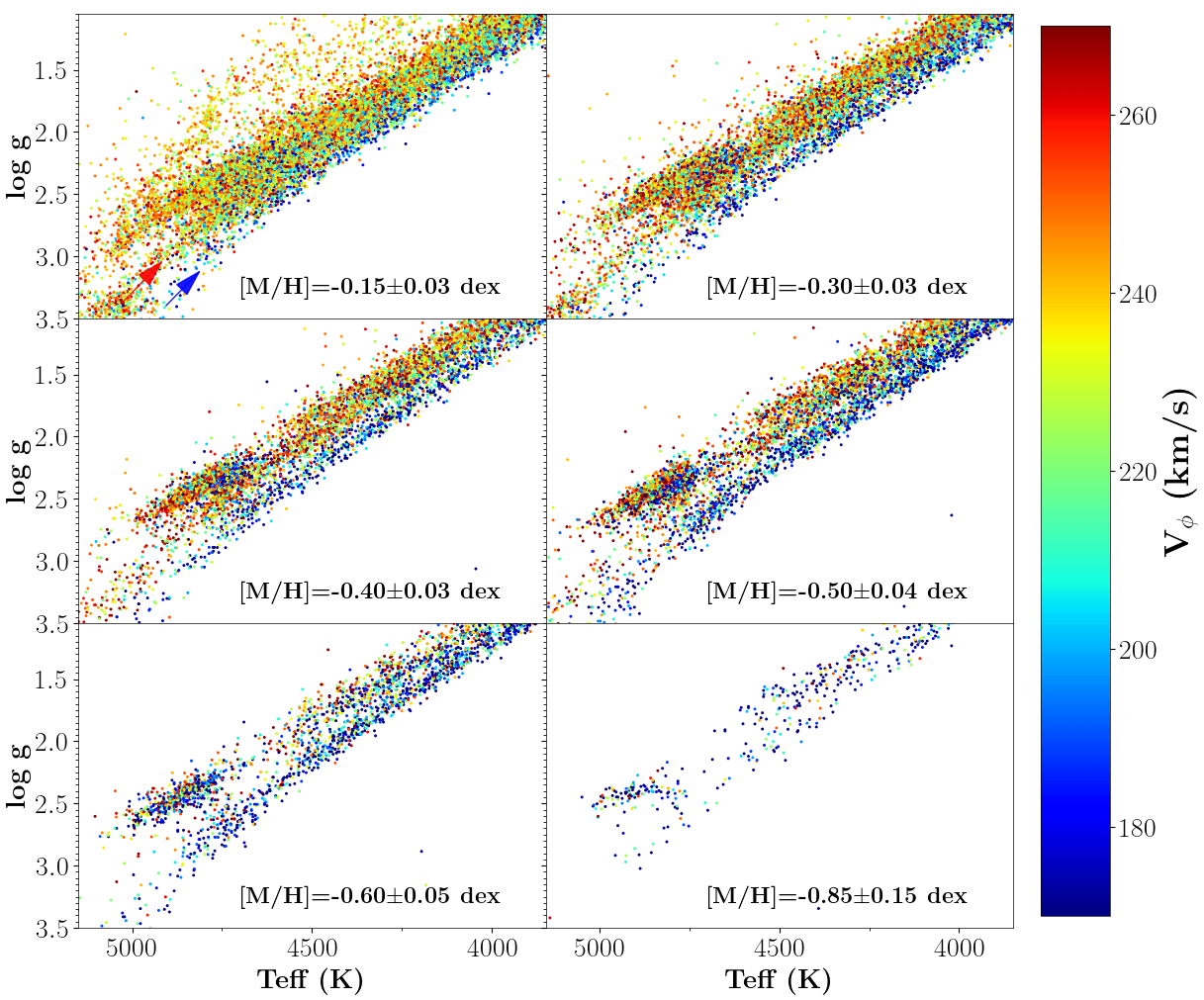} 
\caption{Giant branch and RC locus of the \T\ vs \g\ diagram in six different metallicity bins. The colour code illustrates the azimuthal Galactic velocity. Two RGBs and two RCs are distinguishable in all the panels, with the exception of the more metal-poor one at \meta$=$-0.85~dex, in which only lower Galactic rotational velocity stars are present.}
\label{fig:doubleGiantBranchesVphi}
\end{figure*}

\section{Bimodal giant branches and red clumps }
\label{sect:DoubleRGBs}


\begin{figure*}[h]
\centering
\includegraphics[width=0.8\textwidth]{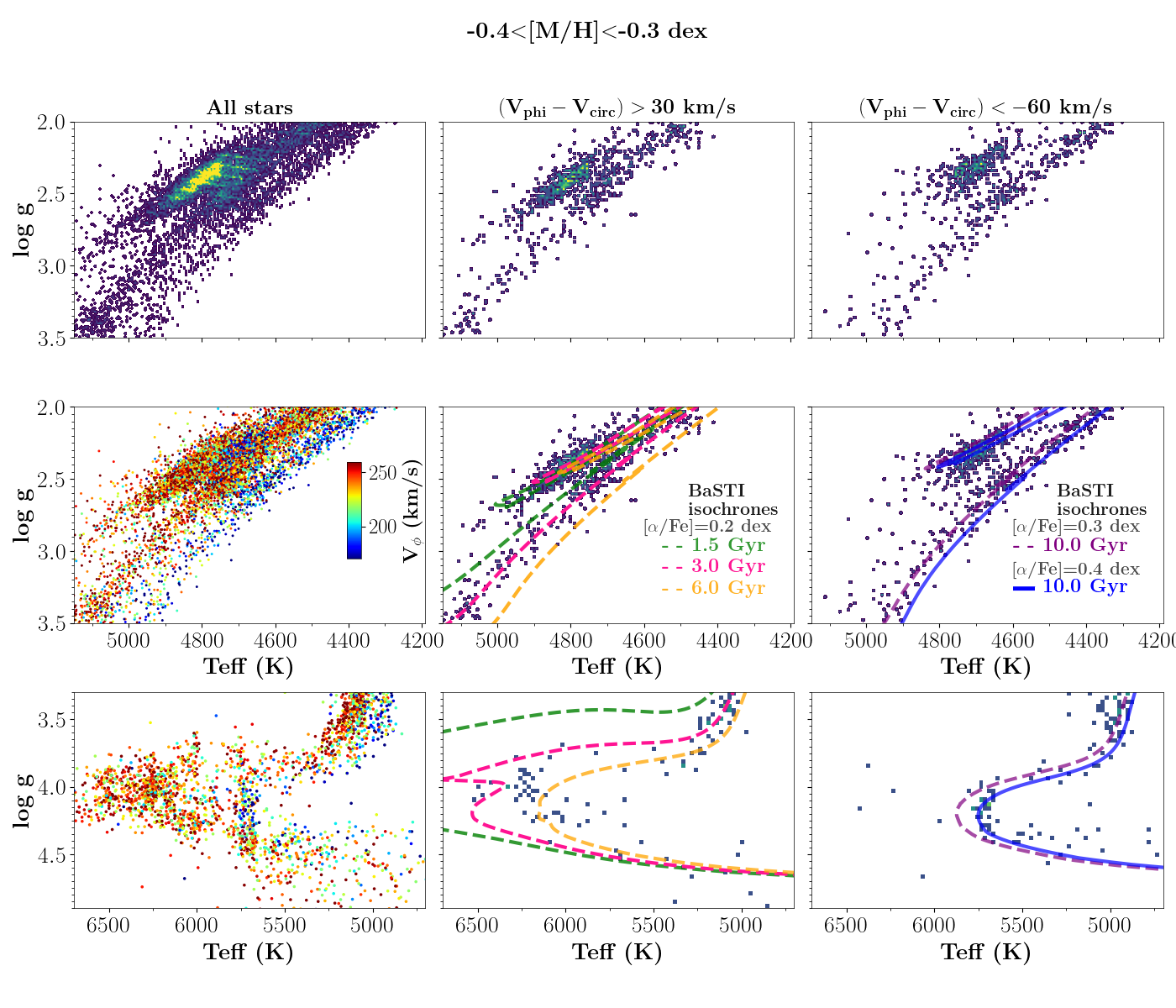} 
\caption{Disc stellar population with \meta$=$-0.35$\pm$0.05~dex. The panels show a kinematical decomposition of the RGB and RC (first and second row) and the turn-off (bottom row) features. The left panels show all the selected stars and present the double evolutionary sequences both for giants and dwarfs. The central and right panels show the thin and thick disc populations, respectively, separated thanks to their $V_\phi$. BaSTI isochrones with different $\alpha$-enhancements are fitted to the data . 
}
\label{fig:MetalIntermediateBin}
\end{figure*}

\begin{figure}[t]
\centering
\includegraphics[width=0.5\textwidth]{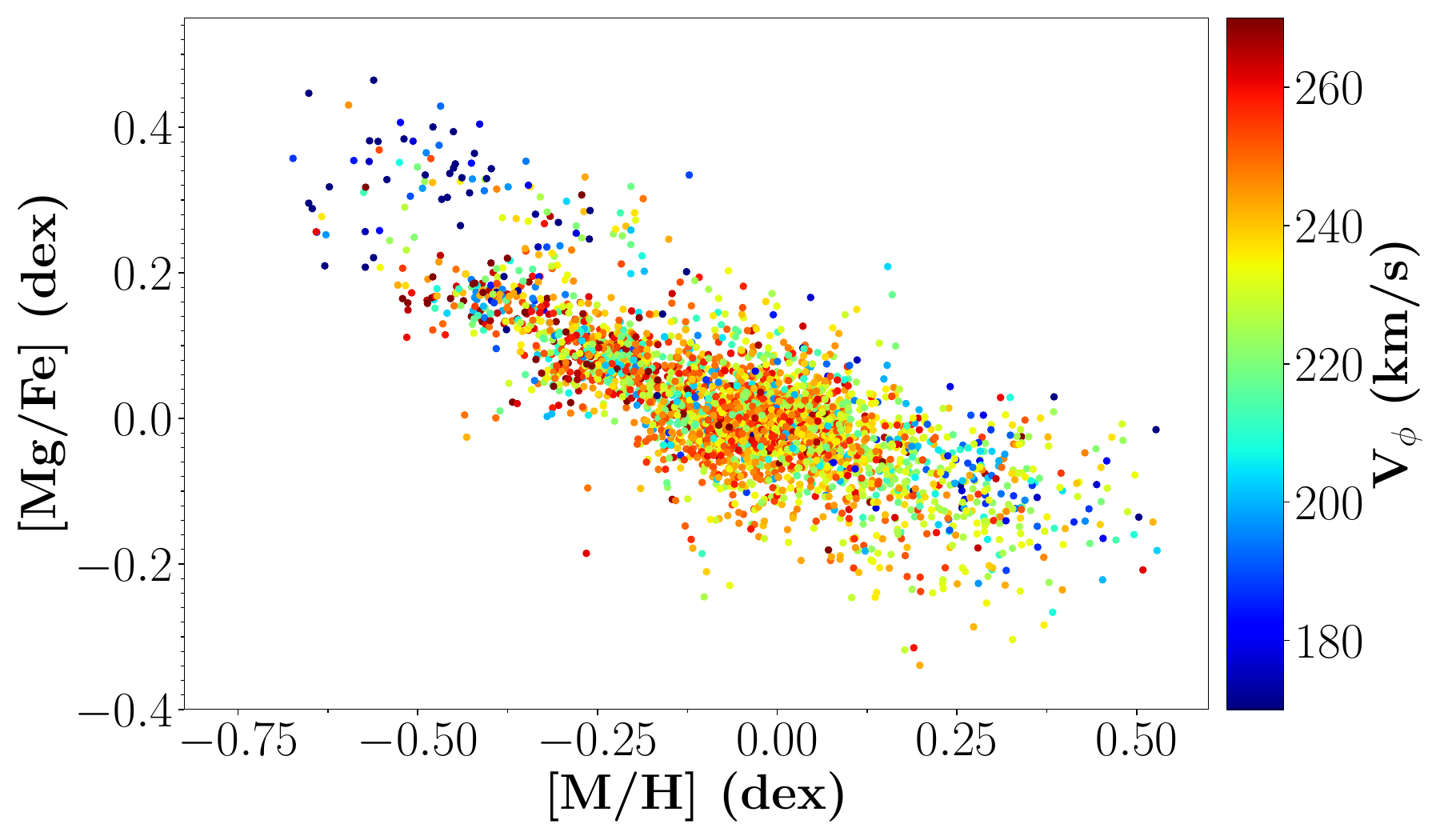}
\caption{\textit{Gaia} \gspspec\ \MgFe\ abundances as a function of metallicity for a sub-sample of 3,209 stars with high S/N data, good quality flags, and uncertainties in \MgFe\ lower than 0.05~dex. The colour code shows the Galactic azimuthal velocity. 
}
\label{fig:MgFeDichotomy}
\end{figure}

\begin{figure}[h]
\centering
 \includegraphics[width=0.5\textwidth]{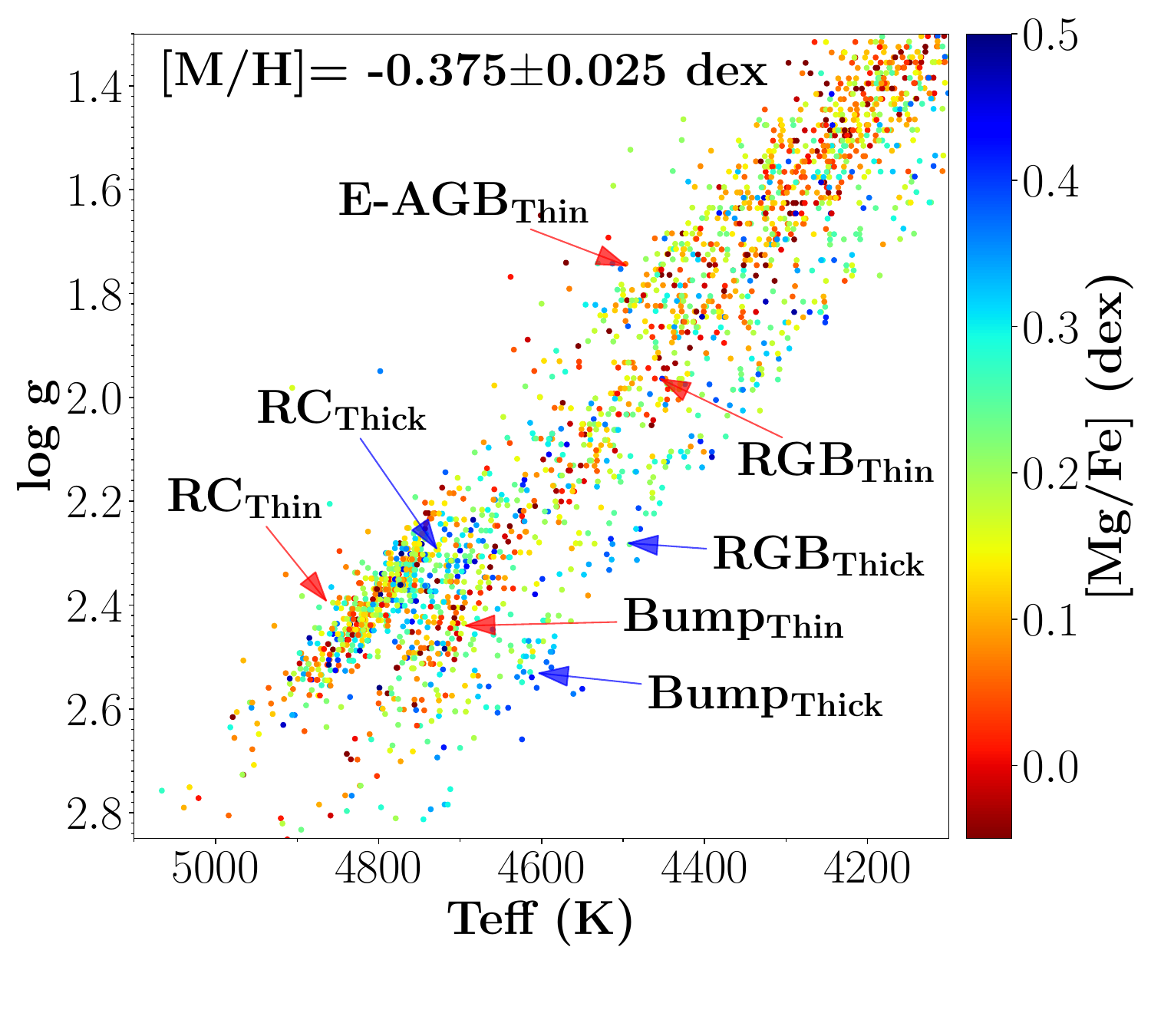} 
\caption{\T\ vs \g\ diagram {for 6\,601 stars} with \meta$=$-0.375$\pm$0.025~dex, using the \MgFe\ abundance as a colour code. 
The main stellar evolutionary features of thin and thick disc populations can be identified.}
\label{fig:KielMgFe}
\end{figure}

\begin{figure}[h]
\centering
\includegraphics[width=0.5\textwidth]{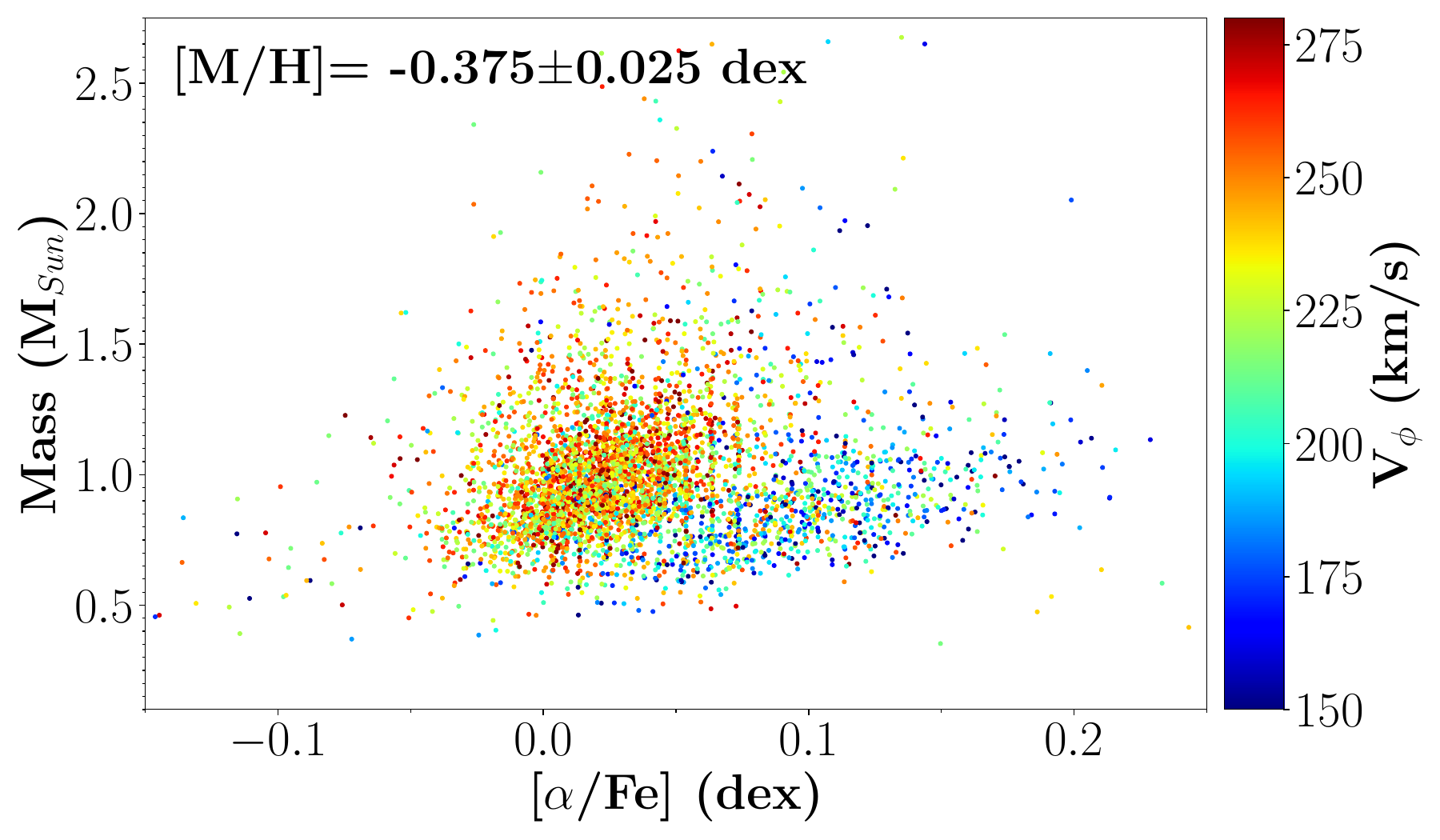}
\caption{Stellar mass as a function of \alphaFe\ for 6\,601 stars with \meta$=$-0.375$\pm$0.025~dex, using the Galactic azimuthal velocity as colour code. Thick disc stars, with lower $V_\phi$ velocities, lower masses, and higher \alphaFe, define a mass-\alphaFe\ sequence separated from thin disc stars.}
\label{fig:Mass}
\end{figure}

\subsection{Kiel diagram perspective}

Thanks to the high number statistics of the \textit{Gaia} \gspspec\ catalogue, our high-precision working sample can be used to confidently select stars in narrow metallicity bins. For each bin, the upper and lower confidence values of the \gspspec\  metallicity indeed allow us to build stellar sub-samples whose \meta\ is within that bin with a probability of 70\%.   This metallicity selection is presented in Fig.~\ref{fig:doubleGiantBranchesVphi}, again using a colour code on the Galactic azimuthal velocity, V$_\phi$. The surface gravity versus temperature distribution of giant stars is shown, considering six separate \meta\ bins. Two different stellar evolutionary paths, highlighted by a red and a blue arrow in the upper left panel, are observed. They are separated by a gap in the \T\ versus \g\ plane, and they are visible from the RGB and the RC up to the asymptotic giant branch. 

It can be seen in this Fig.~\ref{fig:doubleGiantBranchesVphi} that a first evolutionary path is defined by stars with high Galactic rotational velocities, appearing in reddish colours. This population progressively fades from \meta$=$-0.15~dex (upper left panel) to \meta$=$-0.60~dex (bottom left panel), being absent in the more metal-poor bin at \meta$=$-0.85~dex (bottom right panel). 

The second evolutionary path is visible on the cooler side of the first one and is populated by stars with lower Galactic rotational velocities (bluish colours in the figure). It progressively detaches from the high V$_\phi$ evolutionary path as metallicity diminishes, showing a separated RGB in the metal-intermediate regime. This low-V$_\phi$ population is increasingly visible as metallicity decreases, and is the only one present in the most metal-poor bin.

The gap in \T\ and/or \g\ between the two evolutionary paths of mono-metallicity populations suggests different age distributions, with the hotter population being younger than the other one. This fact, combined with the kinematical characteristics (high Galactic rotation for the hotter population and mild Galactic rotation for the cooler one) advocates for their identification as the evolutionary loci of thin and thick disc stars, respectively. This is also supported by the dynamical characteristics of the two populations, as is shown in Fig.\ref{fig:doubleGiantBranchesRperi} and \ref{fig:doubleGiantBranchesZmax}. The cooler, high-V$_\phi$ population is observed to have orbits with higher pericentres (R$_{peri}$) and lower maximum distances from the Galactic plane (Z$_{max}$) than the hotter, lower V$_\phi$ population at a given metallicity. This is coherent with the expected orbital characteristics of the thin disc; namely, a longer scale-length and a shorter scale-height with respect to the thick disc.

To further constrain the characteristics of the observed thin and thick disc bimodality in the  \T\ versus \g\  plane, Fig.~\ref{fig:MetalIntermediateBin} presents a more detailed analysis of an intermediate metallicity bin selecting stars with \verb|mh_gspspec_lower|$>$-0.4~dex and \verb|mh_gspspec_upper|$<$-0.3~dex. The figure is composed of nine panels zooming in on different regions of the Kiel diagram and kinematically decomposing the thin and thick disc evolutionary sequences. The upper and middle row panels show the region around the RGB and the RC, while the main-sequence turn-off region is presented in the bottom row panels. In addition, a kinematical decomposition is displayed, from left to right. No kinematical selection is made in the leftmost panels, showing all the stars in this metallicity bin. On the contrary,  the middle and right column panels present the stars with thin disc or thick disc kinematics, respectively. The kinematical selection is based on the departures from the circular velocity value at the corresponding Galactic position of the star. In particular, thin disc stars in the middle panels are required to have a V$_\phi$ higher than 30 km/s with respect to their circular velocity, V$_{circ}$. The thick disc population in the right panels is selected to have a V$_\phi$ lower than -60 km/s with respect to V$_{circ}$ \citep[see for instance][]{Nadege21}. These criteria avoid as much as possible the overlapping regime in  V$_\phi$ between the thin and thick discs. This allows us to cleanly separate the evolutionary sequences in the Kiel diagram using an independent parameter. In the different panels of Fig.~\ref{fig:MetalIntermediateBin}, two RGBs and two turn-offs separated by a small gap can be identified. In addition, two RCs shifted in both \T\ and \g\ are observed. 

Contrary to the colour-magnitude diagram, the spectroscopic \T\ versus \g\  plane has the advantage of being reddening-free, which is particularly important for disc stars outside the solar neighbourhood. 
As a consequence, and thanks to the \gspspec\ high precision in the stellar parameters of the selected stars, the age distribution of the observed two populations can be confidently retrieved through isochrone fitting. 
For this purpose, we have adopted the BaSTI-IAC isochrone library for both solar-scaled \citep{BaSTI18} and $\alpha$-enhanced \citep{BaSTI21} heavy element distribution \footnote{Since the BaSTI models are available for [$\alpha$/Fe]=0.00 and +0.40, at any given [Fe/H] and age a linear interpolation in [$\alpha/Fe$] has been adopted in order to derive isochrones for any requested $\alpha$-element enhancement. Only a 45~K offset seems to be needed to match the BaSTI isocrones and the calibrated \gspspec\ data, confirming the coherence of both \T\ scales.} and \FeH$=$-0.35~dex (the central value of the analysed metallicity bin). On the one hand, as is shown in the central panel of Fig.~\ref{fig:MetalIntermediateBin}, the thin disc RGB and RC at this metallicity can be fitted with isochrones in the range 1.5 to 6 Gyr. This age distribution is compatible with the isochrone fit of the turn-off populations, presented in the middle bottom panel. 
On the other hand, the thick disc RGB and RC follow the loci of a $\sim$10 Gyr isochrone, as is presented in the middle right panel. This value is again compatible with the fit of the turn-off population in the bottom right  panel. The thick disc evolutionary sequences seem to present a lower age dispersion than the thin disc ones, as is expected from the tighter age-metallicity correlation of the thick disc \citep[e.g.][]{MishaHarps}. Moreover, the fitted thick disc isochrones illustrate the effect of the \alphaFe\ enrichment for the same metallicity and age. Although a shift in \T\ and \g\ is visible and should be considered for careful age determinations on a star-by-star basis,  it is insufficient to explain the gap between the thin and thick disc evolutionary sequences at this metallicity. An age gap of about 4 to 7~Gyr is required to interpret the observed dichotomy.   It should be noted that deriving ages for individual stars is, although possible, beyond the scope of the present paper. Isochrone fitting is employed solely for the purpose of assessing the age range of the observed evolutionary features. 

\subsection{Alpha element abundance and stellar mass}
A common way of separating the stellar populations in the thin and the thick discs is to use the distribution of individual abundances in the \alphaFe\ versus \FeH\ plane. Nevertheless, the separation power of this chemical diagnostic depends on the considered $\alpha$-element\footnote{Those elements whose production is dominated by Type~II supernovae, as Mg, show a higher difference in their abundance with respect to iron between thin and thick disc stars at constant metallicity. 
The \textit{Gaia} \gspspec\ \alphaFe\ estimation is dominated by the Ca triplet \citep{GSPspec2023} and it is therefore a tracer of \CaFe, which has a lower separating power between the thin and thick disc than \MgFe} \citep[e.g.][]{Sarunas2014,Nikos2023}. 
Figure~\ref{fig:MgFeDichotomy} presents the \MgFe\ abundance distribution as a function of metallicity for a sub-sample of stars of our working dataset with high-quality magnesium abundances. In particular, we have selected stars whose uncertainty in \MgFe\ is lower than 0.05~dex and the two corresponding Mg \gspspec\ quality flags are equal to zero. In addition, to avoid potential problems due to line blending,  a \T\ threshold ranging between 4500~K in the metal-poor regime and 5200~K in the metal-rich one has been imposed. As was expected, two sequences are visible in the resulting abundance distribution presented in Fig.~\ref{fig:MgFeDichotomy}. To help with the identification of thin and thick disc populations, corresponding to the lower and the upper sequences, respectively, the colour code shows the Galactic azimuthal velocity of the stars. It is worth noting that the low-V$_{\phi}$ stars visible in the metal-rich regime have orbital pericentres lower than about 6.5~Kpc from the Galactic centre, consistent with their thick disc kinematics and \MgFe\ abundance.

The chemical bimodality can be used to confirm the interpretation of the double giant star sequences presented above. To this purpose, we focus on a very narrow metallicity bin composed of 6,601 stars with \meta$=$-0.375$\pm$0.025~dex. Figure~\ref{fig:KielMgFe} presents the \T\ versus \g\ diagram centred on the giant stars' locus and using the \MgFe\ abundance as a colour code. The bimodality of disc populations is present in all the different stellar evolutionary sequences along this diagram. 
As was expected (cf. Fig.~\ref{fig:MgFeDichotomy}), the thin and the thick disc evolutionary sequences correspond to two different regimes of \MgFe\ abundance, with thick disc stars presenting higher values. In particular, at a metallicity of -0.375~dex, the thick and thin disc sequences in Fig.\ref{fig:MgFeDichotomy} present a [Mg/Fe] abundance of about +0.32~dex and +0.15,~respectively, with a dispersion of about $\pm$0.05~dex in both cases. This corresponds to the observed [Mg/Fe] abundance values for the two RGB sequences in Fig. \ref{fig:KielMgFe}. 

 Finally, Fig.~\ref{fig:Mass} presents the derived stellar masses of the stars in the above-mentioned metallicity bin (\meta$=$-0.375$\pm$0.025~dex)  as a function of their \alphaFe\ abundance.  The selected stars have an uncertainty lower than 0.05~M$_{\sun}$ in mass and 0.01~dex in  \alphaFe. The Galactic azimuthal velocity is used as a colour code. The thin and thick disc stars again define two different regimes in this plot. In particular, the population of stars typical of thick disc stars ( with 
V$_{\phi}$ values lower than 200~km/s)  has a median mass 0.1~M$_{\sun}$ lower than stars with higher V$_{\phi}$ values in the thin disc. This is consistent with the age difference between the two populations. 

\section{Solar and super-solar metallicity disc populations}
\label{sect:VeryMetalRich}

\begin{figure}[h]
\centering
\includegraphics[width=0.5\textwidth]{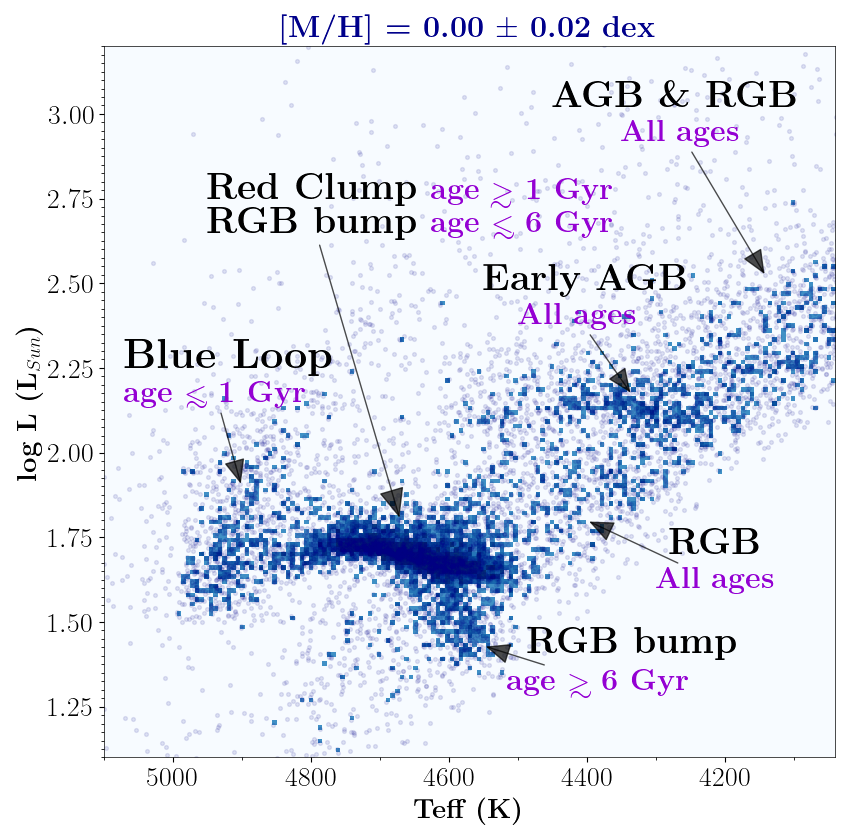} 
\caption{Luminosity vs effective temperature distribution of 12\,532 solar metallicity giant stars selected in a narrow metallicity bin (\meta$=$0.00$\pm$0.02 dex). The main stellar evolutionary features and their corresponding approximate age intervals are presented.}
\label{fig:SolarMetallicity}
\end{figure}

As is shown in Fig.\ref{fig:doubleGiantBranchesVphi}, the thin and thick disc RGB and RC features become closer as metallicity increases. Around solar metallicity, the evolutionary sequences merge as a consequence of the smaller age and \alphaFe\ differences between the two disc populations in this regime. Figure~\ref{fig:SolarMetallicity} presents the luminosity versus \T\ distribution of 12\,532 solar metallicity giant stars selected in the narrow metallicity bin \meta$=$0.00$\pm$0.02 dex. The morphology of the resulting diagram reflects the principal evolutionary stages of the stars in this bin, as is illustrated in the figure. The possibility of dissecting the stellar populations in this kind of mono-abundance population plot opens up new avenues for Galactic studies. For instance, blue loop stars were selected from \gspspec\ Kiel diagrams to trace the young stellar populations in the spiral arms  by \cite{PVP2023} and \cite{Eloisa22}.

\begin{figure}[h]
\centering
\includegraphics[width=0.5\textwidth]{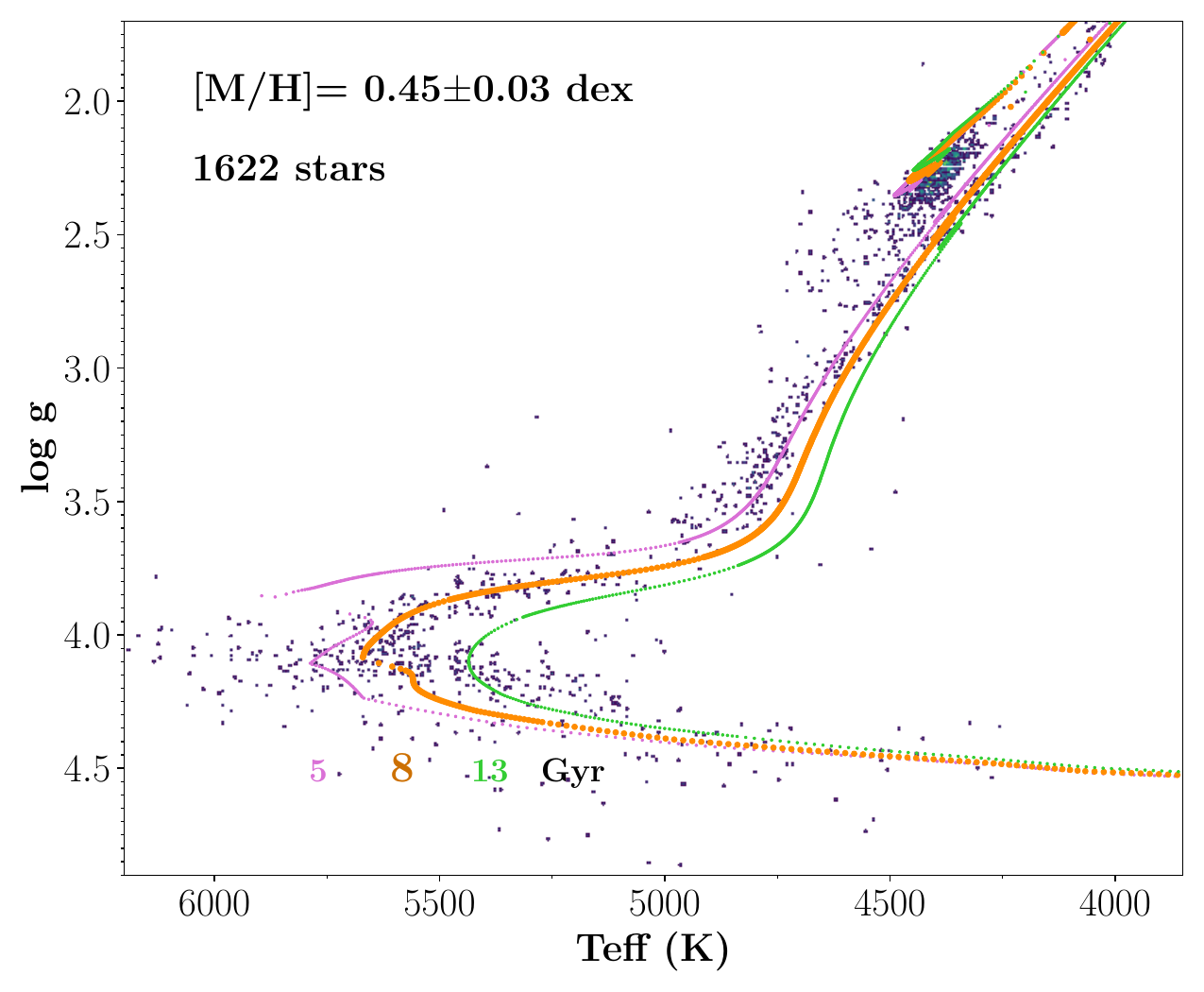} 
\caption{Kiel diagram of the extremely metal-rich disc population with  \meta$=$+0.45$\pm$0.03 dex. Solar-scaled BaSTI isochrones of 5, 8, and 13 Gyr (pink, orange, and green, respectively) are overlaid on the data.}
\label{fig:SuperMetalRich}
\end{figure}

Moreover, the study of disc stars presenting the highest levels of metal enrichment is key to understand the mechanisms of radial migration \citep[e.g.]{SellwoodBinney,RalphS}. Disc super-metal-rich stars have been the subject of extensive analysis \citep[see for instance][]{Grenon99, Trevisan11, Dantas23, Nepal24}, with a considerable body of evidence pointing to their origin in the more internal regions of the disc.  
Figure~\ref{fig:SuperMetalRich} shows the Kiel diagram for the metallicity bin \meta=+0.45$\pm$0.03~dex (a factor $\sim$3 larger than the solar metal enrichment), and the corresponding fit of three BaSTI isochrones with 5, 8, and 13 Gyr. An important fraction of the stars in this extremely metal-rich regime are old. The different evolutionary features are globally well fitted by a 8~Gyr isochrone, although younger and older isochrones in the interval of about 5 to 13 Gyr are also compatible with about $85\%$ of the selected stars. Interestingly enough, this age interval is similar to the one observed for bulge populations by \cite{Joyce23} and compatible with previous studies of super-metal-rich populations. 
Additionally, about 15$\%$ of the stars are in the temperature regime of younger populations, with \T\ values between 6000 and 8000~K, and possibly on the hotter side of the RC. Although the relative proportion of young to old stars is difficult to evaluate (among other things, due to a lack of completeness), the data highlight the existence of an old extremely metal-rich population. It is interesting to  note that the radial distribution of stars in this metallicity bin spans from 7.5 to 9~kpc from the Galactic centre (with a median at 8.24~kpc). The distribution of maximum orbital distances from the Galactic plane, Z$_{max}$, ranges from  0 to 1~kpc (with a median at 0.26~kpc). Finally, the median Galactic azimuthal velocity is 235~km/s, with a dispersion of 25~km/s.

\section{Conclusions}
\label{sect:Conclu}
The Gaia DR3 \gspspec\ catalogue allows one to study the disc stellar populations in a high-precision chemo-physical parameter regime, with unprecedentedly high number statistics. This opens new horizons to characterise in detail the thin and thick discs' evolution outside the solar neighbourhood, getting rid of interstellar absorption issues and breaking the age-metallicity degeneracy. In this work, we present for the first time the consequence of Galactic disc bimodality in the physical properties of its giant star population; that is, double RGB sequences and RC features for mono-metallicity populations. An age gap is needed to explain the evolutionary sequence separation, in agreement with previous age-metallicity relations using sub-giant stars \citep[e.g.][]{XiangRix2022} and supporting the modelling of the disc chemical evolution by  distinct infall episodes \citep[e.g.][]{Emanuele19, EmanueleDR3}. The bimodal evolutionary sequences are characterised in kinematics, dynamics, \MgFe\ abundances, and stellar masses consistently derived from Gaia data. The thin and thick disc RGB and RC features become closer as metallicity increases, merging around solar metallicity as the result of the smaller age and \alphaFe\ differences between the two disc populations in the metal-rich regime. A selection of extremely metal-rich disc stars with \meta=0.45$\pm$0.03~dex contains a considerable proportion of very old stars (ages 5-13 Gyr) reaching distances of up to 9~kpc from the Galactic centre, and maximum orbital distances from the Galactic plane of up to 1~kpc. This old population could be composed of thin disc stars that have migrated from the internal regions of the Galaxy, and thick disc stars formed before the last major merger of the Milky Way; namely, that of the Gaia-Enceladus satellite about 9~Gyrs ago \citep{Amina18, Vasili18, Carme2019}. The detailed analysis of precise Gaia \gspspec\ Kiel diagrams of mono-abundance stellar populations opens new pathways to disentangle the complex puzzle of Galactic disc bimodality that is at the core of Milky Way's evolutionary processes.

\begin{acknowledgements}
This work has made use of data from the European Space Agency (ESA)
mission Gaia (https://www.cosmos.esa.int/gaia), processed by the Gaia Data Processing and Analysis Consortium (DPAC, https://www.cosmos.
esa.int/web/gaia/dpac/consortium). Funding for the DPAC has been provided by national institutions, in particular the institutions participating in the Gaia Multilateral Agreement. PAP acknowledges support by the Centre National d'études Spatiales (CNES).  SC and AP acknowledge financial support from PRIN-MIUR-22: CHRONOS: adjusting the clock(s) to unveil the CHRONO-chemo-dynamical Structure of the Galaxy” (PI: S. Cassisi) finanziato dall’Unione Europea – Next Genera- tion EU, and Theory grant INAF 2023 (PI: S. Cassisi). We are grateful to the anonymous referee for their constructive feedback, which has significantly improved the clarity of the manuscript.

\end{acknowledgements}


\bibliographystyle{aa}  
\bibliography{biblio} 

\begin{appendix}

\section{Calibration of surface gravity, metallicity, and magnesium abundance}
\label{App:Calibs}

\begin{figure*}[ht]
\centering
\includegraphics[width=0.95\textwidth]{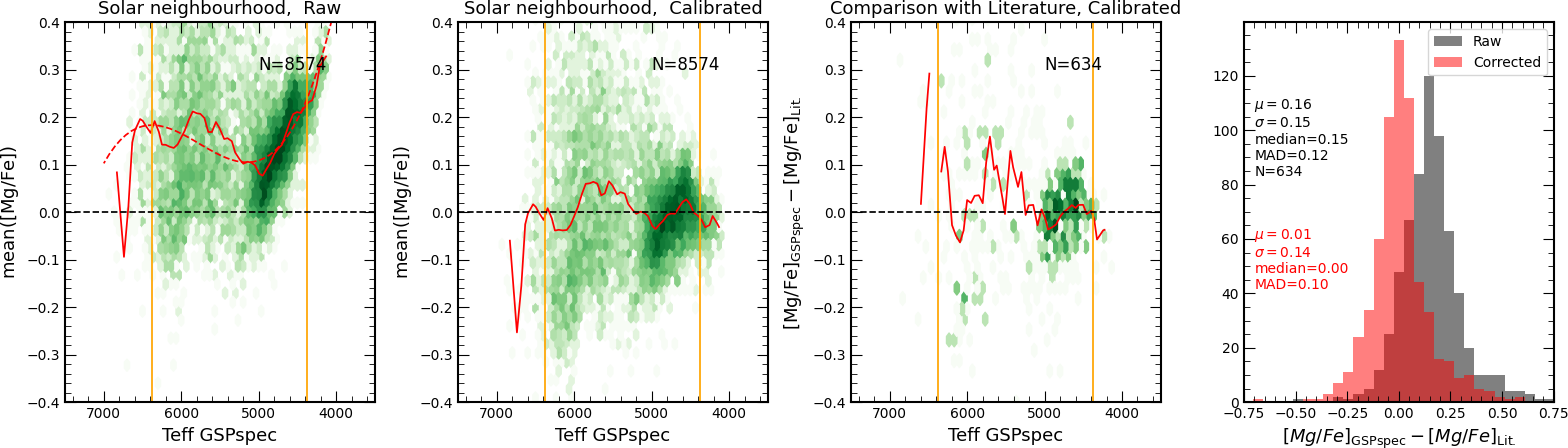} 
\caption{Polynomial fits for the calibration of the \MgFe\ abundance as a function of \T\ following the procedure of \cite{GSPspec2023}. The two rightmost panels present the comparison with literature data before and after the calibration.}
\label{fig:MgFeCalib}
\end{figure*}

The raw \gspspec\ values for surface gravity, metallicity and magnesium abundance with respect to iron have been calibrated following the procedures described in \cite{GSPspec2023}. However, as the temperature range spanned by these data is large, it has been found to be more adapted to calibrate as a function of \T\, instead of as a function of \g. To this purpose, and following \cite{GSPspec2023}, polynomial corrections in the form:

\begin{equation}
Param_{\rm calibrated}=Param+ \sum_{i=0}^3 p_i \cdot t^i
\end{equation}

have been applied, where $Param$ is either \g, \meta\ or \MgFe. $t$ is the relative temperature with respect to the Sun, \T/5750, and $p_i$ are the corresponding polynomial coefficients from Table~\ref{tab:calibrations}. Figure \ref{fig:MgFeCalib} illustrates the calibration data for \MgFe, the corresponding polynomial fits and the comparison with literature abundances.

\begin{table}[h]
\centering
    \caption{Polynomial coefficients for the calibration of the \gspspec\ MatisseGauguin gravities, metallicities and \MgFe\ abundances as a function of $t=$\T/5750}
    \label{tab:calibrations}
    \begin{tabular}{l|cccc}
        \hline
        Parameter & $p_0$ & $p_1$ & $p_2$ & $p_3$  \\ 
        \hline
         \g & -0.2320 & 0.8291 & 0.7441 & -1.1677 \\
         \meta & 1.2739 & -1.1383 & -1.4776 & 1.3398  \\
         \MgFe & -15.8586 & 48.1211 & -48.5437  & 16.1383 \\
        \hline
    \end{tabular}
\end{table}

\section{ADQL query to the \textit{Gaia} archive} 
\label{App:Query}
The \textit{Gaia} \gspspec\ data sample used in this work can be retrieved from the \textit{Gaia} archive with the following query:
\lstset{language=SQL}

\begin{lstlisting}[caption={\texttt{ADQL} query of the stars with GSP-Spec parameters used in this work.},captionpos=b]
SELECT source_id
FROM gaiadr3.astrophysical_parameters LEFT JOIN gaiadr3.gaia_source USING(source_id)
WHERE ((mh_gspspec_upper-mh_gspspec_lower)<0.05) AND (teff_gspspec>3750) AND (logchisq_gspspec<-3.7) AND (flags_gspspec LIKE '000000%') AND ((flags_gspspec LIKE '_______0%') OR (flags_gspspec LIKE '_______1%')) AND (flags_gspspec LIKE '____________0%') AND (rv_expected_sig_to_noise>100)

\end{lstlisting}

\section{Density diagrams and dynamical dependencies}
\label{App:MorePlots}
\begin{figure}
\centering
\includegraphics[width=0.5\textwidth]{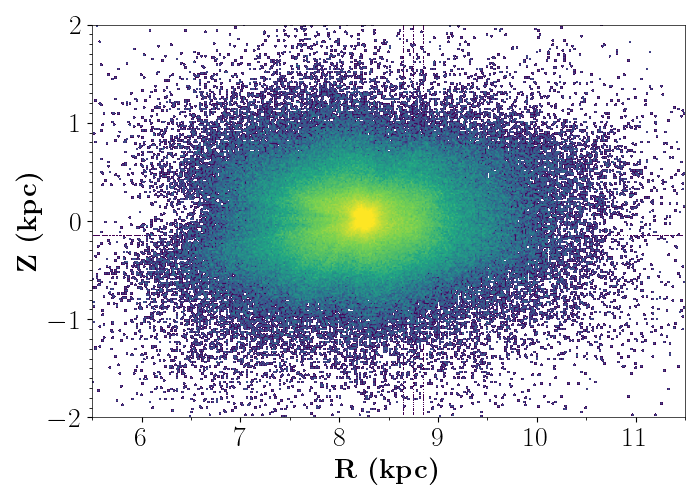} 
\caption{ Density plot showing the spatial distribution of the $\sim$408,800 stars in the \gspspec\ very high precision sample, used in the present analysis. }
\label{fig:RZ}
\end{figure}

\begin{figure*}
\centering
\includegraphics[width=0.65\textwidth]{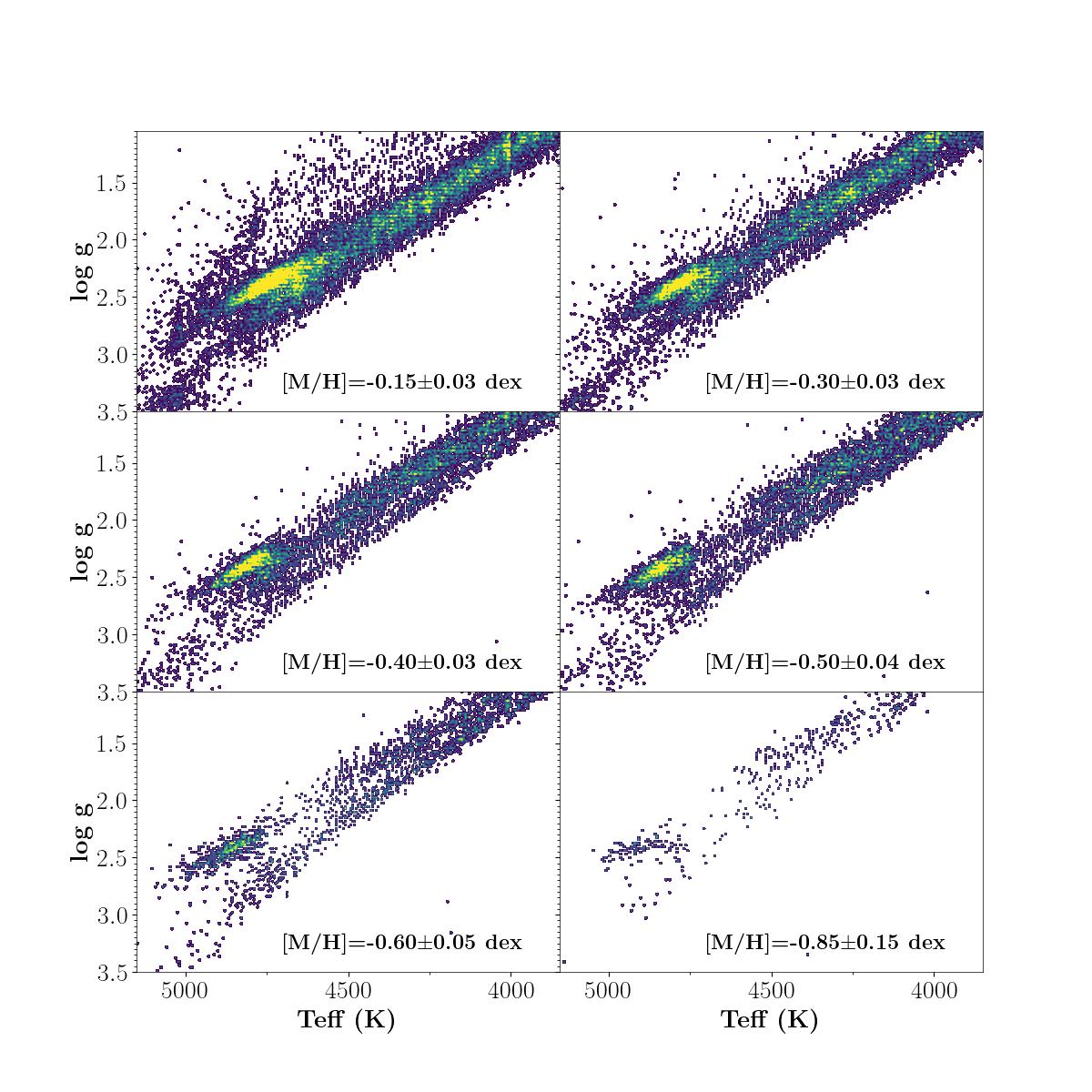} 
\caption{Density plot corresponding to Fig.~\ref{fig:doubleGiantBranchesVphi} stellar populations.}
\label{fig:doubleGiantBranches}
\end{figure*}

\begin{figure*}
\centering
\includegraphics[width=0.65\textwidth]{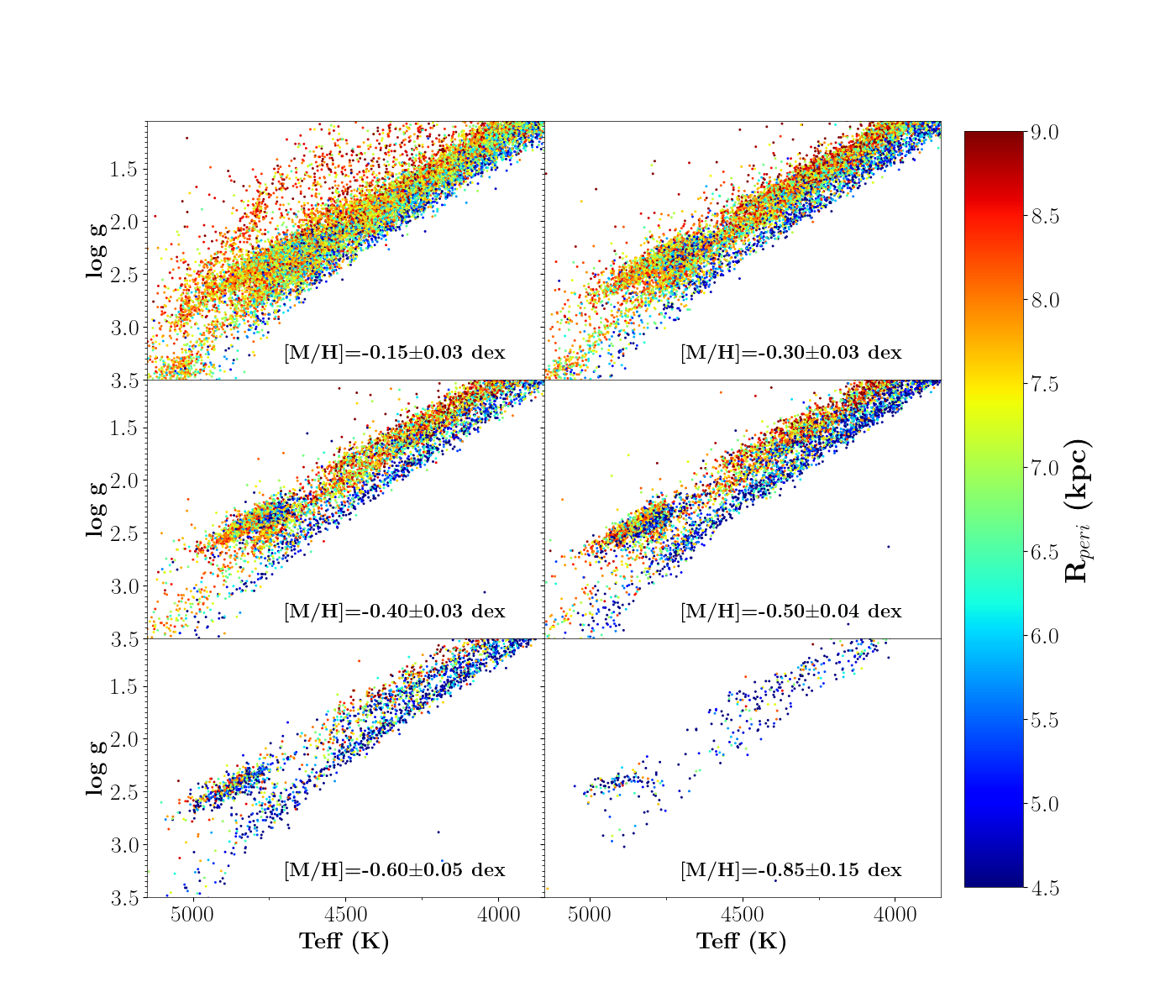} 
\caption{Same as Fig.~\ref{fig:doubleGiantBranchesVphi}, but colour-coded with the star orbital pericentre (R$_{peri}$).}
\label{fig:doubleGiantBranchesRperi}
\end{figure*}

\begin{figure*}
\centering
\includegraphics[width=0.65\textwidth]{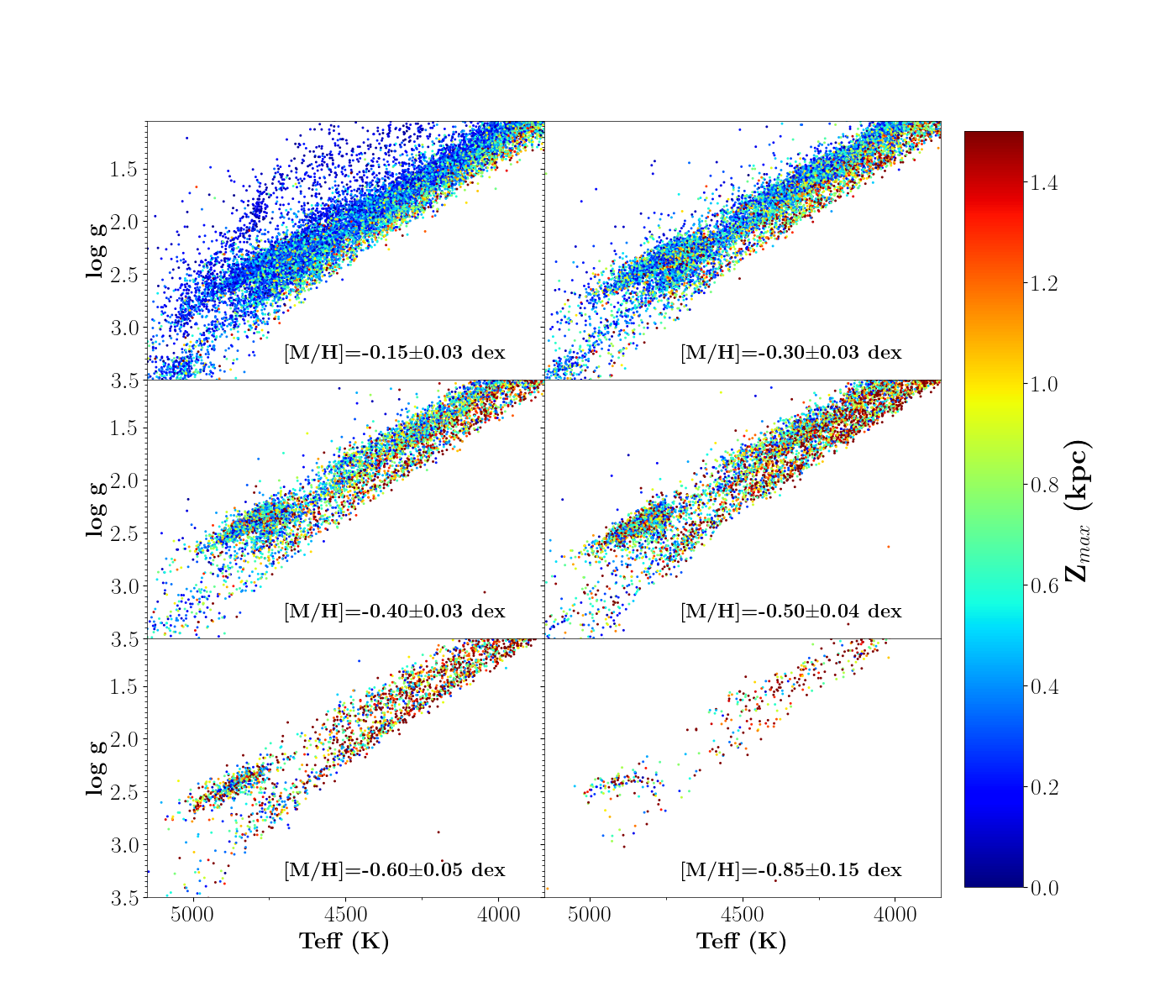} 
\caption{Same as Fig.~\ref{fig:doubleGiantBranchesVphi}, but colour-coded with star maximum orbital distance to the Galactic plane (Z$_{max}$), as a colour code.}
\label{fig:doubleGiantBranchesZmax}
\end{figure*}

Figure~\ref{fig:RZ} presents the spatial distribution of the $\sim$ 408,800 stars with very high precision \gspspec\ parameters used in this article, clearly covering the thin and thick disc Galactic locii.

The thin/thick disc bimodality in the Kiel diagram illustrated for six different mono-abundance populations in Sect.~\ref{sect:DoubleRGBs} is also distinguishable in the stellar distribution in this plane (see Fig~\ref{fig:doubleGiantBranches}). Thanks to the precision of the \gspspec\ chemo-physical parameters, the double RGB sequences are visible without the use of any kinematical or dynamical information, particularly in the intermediate metallicity domain. 

In addition, Figs.~\ref{fig:doubleGiantBranchesRperi} and \ref{fig:doubleGiantBranchesZmax} allow to observe the trends of the two RGB sequences with the orbital pericentre, R$_{peri}$, and the maximum distance from the Galactic plane, Z$_{max}$, respectively. The typical thin and thick disc ranges in R$_{peri}$ and  Z$_{max}$ can be distinguished. The hotter sequence presents larger orbital pericentres and lower Z$_{max}$ values, typical of thin disc stars. On the contrary, the cooler sequence corresponds to a population with pericentres closer to the Galactic centre and reaching larger distances from the plane, characterising thick disc stars.


\end{appendix}

\end{document}